\documentclass[12pt,a4,portrait]{article}
\usepackage{latexsym}
\usepackage{amssymb}
\usepackage{amsmath}
\usepackage{hyperref}
\usepackage{graphicx}

\newcommand{\be}{\begin{equation}}
\newcommand{\ee}{\end{equation}}
\newcommand{\beq}{\begin{eqnarray}}
\newcommand{\eeq}{\end{eqnarray}}
\newcommand{\bed}{\begin{displaymath}}
\newcommand{\eed}{\end{displaymath}}
\newcommand{\bc}{\begin{center}}
\newcommand{\ec}{\end{center}}
\newcommand{\bi}{\begin{itemize}}
\newcommand{\ei}{\end{itemize}}
\newcommand{\bn}{\begin{enumerate}}
\newcommand{\en}{\end{enumerate}}

\newcommand{\rw}{ {\rm w} }

\begin{document}

\title{Scalar-tensor cosmologies with dust matter \\ in the general relativity limit}
\author{Laur J\"arv\thanks{laur.jarv@ut.ee}, Piret Kuusk\thanks{piret@fi.tartu.ee},
Margus Saal\thanks{margus@fi.tartu.ee} \\
 Institute of Physics, University of Tartu, Riia 142, Tartu 51014, Estonia} 
 
\date{}
\maketitle

\begin{abstract}
We consider flat Friedmann-Lema\^{\i}tre-Robertson-Walker
cosmological models in the framework of general scalar-tensor theories of gravity 
with arbitrary coupling functions, set in the Jordan frame, in the cosmological epoch
when the energy density of the ordinary dust matter dominates over the energy density of
the scalar potential.
Motivated by cosmological observations, we apply an approximation scheme 
in the regime close to the so-called limit of general relativity. 
The ensuing nonlinear approximate equations for the scalar field and the 
Hubble parameter can be solved analytically in 
cosmological time. 
This allows us to distinguish the theories with solutions that asymptotically converge to general relativity and draw some 
implications about the cosmological dynamics near this limit.  

\end{abstract}

Pacs: \  04.50 Kd, 98.80.Jk, 95.36+x

\section{Introduction}

Various cosmological observations of our Universe can be fairly well accommodated within
the $\Lambda$CDM concordance model \cite{LambdaCDM} based on the theory of general relativity (GR).
However, there is still a number of viable alternative theories 
which also manage to conform sufficiently
well with observational data \cite{alternatives_reviews}. One such family of theories 
is provided by scalar-tensor gravity (STG) \cite{books} where gravitational interaction is mediated by an extra scalar degree of freedom $\Psi$ in addition to the usual
tensor ones. In the so-called Jordan frame and Brans-Dicke like parametrization an STG is characterized by
two arbitrary functions, the coupling function $\omega(\Psi)$ and the scalar
potential $V(\Psi)$.
As has been discussed by many authors previously \cite{dn,attractor}, 
for a range of choices of
$\omega$ and $V$ the cosmological evolution of dust and potential dominated STG models naturally converges close to the one expected from GR.
Yet, at the same time STG models may also offer a possibility to explain 
small observational differences from pure GR $\Lambda$CDM behavior, e.g.
the possibly variable effective barotropic index of dark energy
\cite{STG_w} as hinted by some observational data \cite{muutuv_w}, deviations in the growth of perturbations \cite{perturbations}, etc.

The aim of the current paper is to narrow down the class of STG models that
can lead to observationally viable cosmologies (i.e., spontaneously evolve close to GR), 
and by explicitly finding the general solutions applicable in this regime to provide 
a basis for further direct checks with observational data.
It is a follow-up work to
our recent papers \cite{meie5, meie7, meie8} where we investigated Friedmann-Lema\^{\i}tre-Robertson-Walker
(FLRW) cosmological models in the framework of general STG
with arbitrary coupling function and scalar potential in the era when 
the energy density of the scalar potential dominates over the energy density of ordinary matter.
There we presented, justified and applied an approximation scheme for 
the scalar field equation to capture the scalar field dynamics near the GR limit. 
In the present paper we supplement these studies with analogous 
investigations for the the cosmological epoch
when the energy density of the ordinary dust matter dominates over the energy density of
the scalar potential. The presence of an extra dynamical quantity (matter)
in the system makes the procedure now a bit more complicated, yielding
two non-linear equations which explicitly contain time but which can be nevertheless solved analytically.
In a realistic cosmological scenario the dust dominated
epoch should be patched together with the potential dominated era (as well as with an account of the early universe). 
For some related recent studies see Refs. \cite{specific_models}. 
    
In section 2 we recall STG FLRW equations. In section 3 we motivate 
and apply to the dust matter dominated era the approximation method
worked out in Ref. \cite{meie5}.
The resulting nonlinear equations are solved analytically 
in cosmological time in Sect. 4. Comparison with earlier results and implications for selecting a model of STG viable in cosmology are discussed in section 5. 
Finally, section 6 provides a summary and a brief outlook.

\section{The equations of scalar-tensor cosmology}

We consider  a general scalar-tensor theory
in the Jordan frame given by the action functional
    \beq \label{jf4da}
S  = \frac{1}{2 \kappa^2} \int d^4 x \sqrt{-g}
        	        \left[ \Psi R(g) - \frac{\omega (\Psi ) }{\Psi}
        		\nabla^{\rho}\Psi \nabla_{\rho}\Psi  - 2 \kappa^2 V(\Psi) \right]
                   + S_{m}(g_{\mu\nu}, \chi_m) \,.
\eeq 
Here $\omega(\Psi)$ is a coupling function, $\nabla_{\mu}$ 
denotes the covariant derivative with respect to the metric 
$g_{\mu\nu}$, $\kappa^2$ is the non-variable part of
the gravitational constant, and 
$S_{m}$ is the matter contribution to the action 
as all other fields are included in $\chi_m$.
In order to keep the effective gravitational constant $8 \pi G = \frac{\kappa^2}{\Psi}$ positive,
we assume that $0 < \Psi < \infty$.

The field equations for the Friedmann-Lema\^{\i}tre-Robertson-Walker
(FLRW) line element 
\be
ds^2=-dt^2 + a(t)^2 \left( \frac{dr^2}{1-kr^2} + r^2 (d\theta ^2 + \sin ^2 \theta d\varphi ^2)\right)
\ee
with curvature parameter $k= 0$ (flat) and perfect barotropic fluid 
matter, $p=\rw \rho$, $\rw = {\rm const.}$, read 
\beq 
\label{00}
H^2 &=& 
- H \frac{\dot \Psi}{\Psi} 
+ \frac{1}{6} \frac{\dot \Psi^2}{\Psi^2} \ \omega(\Psi)
+ \frac{\kappa^2}{\Psi} \frac{\rho}{3} 
+ \frac{\kappa^2}{\Psi} \frac{V(\Psi)}{3} \,, 
\\ \nonumber \\
\label{mn}
2 \dot{ H} + 3 H^2 &=& 
- 2 H \frac{\dot{\Psi}}{\Psi} 
- \frac{1}{2} \frac{\dot{\Psi}^2}{\Psi^2} \ \omega(\Psi) 
- \frac{\ddot{\Psi}}{\Psi} 
- \frac{\kappa^2}{\Psi} \rw \rho
+ \frac{\kappa^2}{\Psi} \ V(\Psi) \,, 
\\ 
\label{deq}
\ddot \Psi &= & 
- 3H \dot \Psi 
+ \frac{1}{2} \, A(\Psi)(2\omega(\Psi) + 3) \  \dot {\Psi}^2 
+ \frac{\kappa^2}{2 \omega(\Psi) + 3} (1-3\rw)\rho \nonumber\\
&& \qquad + \frac{2 \kappa^2}{2 \omega(\Psi) + 3} \ \left[ 2V(\Psi) - \Psi \ 
\frac{d V(\Psi)}{d \Psi}\right] \, ,
\eeq 
where $H \equiv \dot{a} / a$ and we have introduced the notation
\be
A(\Psi) \equiv \frac{d}{d\Psi} \left(\frac{1}{2\omega(\Psi)+3} \right) 
\ee
for later convenience.
The matter conservation law is the usual 
\beq \label{matter_conservation}
\dot{\rho} + 3 H \ (\rw+1) \ \rho = 0 \,,
\eeq
it is reasonable to assume positive matter density, $\rho \geq 0$. 

The Hubble parameter $H$ can be expressed as a function of $\Psi$ by solving the Friedmann equation (\ref{00}) algebraically,
\beq  \label{H_eq}
H = -\frac{\dot\Psi}{2\Psi} \pm \sqrt{(2\omega(\Psi)+3)\frac{\dot\Psi^2}{12 \Psi^2} + \frac{\kappa^2 (\rho + V(\Psi))}{3 \Psi}} \,.
\eeq
For later argument notice that in the limit  $\frac{1}{(2\omega(\Psi)+3)} \rightarrow 0$, $\dot\Psi \neq 0$ the system 
faces a spacetime curvature singularity, since $H$ diverges. Only as long as $(2\omega(\Psi)+3)\dot\Psi^2$ is 
finite are the solutions singularity free.

Let us take the regime where the dominating contribution to cosmological energy density is provided by dust matter ($\rw=0$) and the 
scalar potential can be neglected in the equations. 
The system (\ref{00})-(\ref{matter_conservation}) is characterized by three variables $\{ \Psi,  H, \rho \}$, 
but one of them is algebraically related to the others via the Friedmann equation (\ref{00}).
Eliminating $\rho$ yields two equations
\beq 
\label{md_ddotpsi}
{\ddot \Psi} &=& - 3H {\dot \Psi} + \frac{1}{2} (2\omega +3) A(\Psi) {\dot \Psi}^2 
+ \frac{1}{(2\omega +3)} \left(3 \Psi H^2 + 3H {\dot \Psi} - \frac{{\dot \Psi}^2}{2 \Psi} \omega \right)\,, \\
{\dot H} &=& -\frac{3}{2} H^2 +  H \frac{{\dot \Psi}}{2\Psi}  - 
 \frac{{\dot \Psi}^2}{4\Psi^2} \omega - \frac{1}{4} (2\omega +3)A(\Psi)\frac{{\dot \Psi}^2}{\Psi} \nonumber \\
  &&- \frac{1}{2(2\omega +3)} \left( 3 H^2 + 3H \frac{{\dot \Psi}}{\Psi} - \frac{{\dot \Psi}^2}{2 \Psi^2} \omega \right)\,,
\label{dotha}
\eeq
which provide the basis for the present study.

\section{Approximate equations}

In fact, not all possible solutions of Eqs. (\ref{md_ddotpsi})-(\ref{dotha}) are of immediate
physical interest, since cosmological observations give a clear 
preference towards a certain corner in the solutions space.
Analysis of the anisotropies of the cosmic microwave background (CMB) radiation
sets a limit on the variation of the gravitational constant
from the recombination process till now, $\frac{|G_{\mathrm{rec}}-G_{\mathrm{now}}|}{G_{\mathrm{now}}}< 5 \times 10^{-2}$ \cite{G_var,Nagata:2003qn}, which for scalar-tensor gravity translates into
$\dot\Psi \ll 1$. In addition, the best fit of the CMB data indicates
that at the time of recombination
$\frac{1}{2 \omega(\Psi_{\mathrm{rec}})+3} < 7 \times 10^{-2}$ 
\cite{Nagata:2003qn}, while the value today from the PPN data is
bounded as
$\frac{1}{2 \omega(\Psi_{\mathrm{now}})+3} < 7 \times 10^{-4}$ \cite{PPN}.

Therefore it makes sense while considering the dust dominated cosmological era to focus upon the solutions near the limit (a) $\frac{1}{(2\omega(\Psi)+3)} \rightarrow 0$ and (b) $\dot\Psi \rightarrow 0$.
This assumption is consistent with the equations, as one can check in Eq. (\ref{deq}) how the conditions (a) and
(b) keep $\ddot{\Psi}$ negligible, thus allowing $\dot{\Psi}$ to remain negligible as well. One may also argue that if changes in $\Psi$ are sufficiently small, $\omega(\Psi)$ does not change dramatically and the regime expected from the solutions is sufficiently stable to merit investigation.

So, let us define $\Psi_{\star}$ by 
\be
\label{Psi_star}
\frac{1}{2\omega(\Psi_{\star})+3} =0
\ee
and focus upon the solutions near this point,
\be \label{xh_def}
\Psi (t) = \Psi_{\star} + x(t) \,, \qquad H(t) = H_{\star}(t) + h(t)\,,
\ee
where $H_{\star}(t)$ is the Hubble parameter corresponding to the cosmological evolution with $\Psi_{\star}$, 
while $x(t)$ and $h(t)$ are small deviations.
It follows from (\ref{xh_def}) that $\dot\Psi (t) = \dot{x}(t)$, where we expect $\dot{x}(t)$ 
to be also small due to (b).
Under the two additional mathematical assumptions, (c) $A_{\star} \equiv A(\Psi_{\star}) \neq 0$ and 
(d) $\frac{1}{2\omega+3}$ is differentiable at $\Psi_{\star}$, 
we can expand in series
\beq
\label{approx_2}
\frac{1}{2\omega(\Psi)+3} &=& \frac{1}{2\omega(\Psi_{\star})+3}+ A_{\star} x+ ...\approx A_{\star} x \,,  \\
(2\omega(\Psi) + 3) \dot{\Psi}^2 &=& \frac{\dot{x}^2}{0 + A_{\star}x + \ldots} =  \frac{\dot{x}^2}{A_{\star} x}  (1 + O(x)) \approx \frac{\dot{x}^2}{A_{\star}x} \,.
\eeq
The latter result actually informs us that in order to avoid a spacetime singularity $\frac{\dot{x}^2}{x}$ must not diverge, hence we should treat $x(t)$ and $\dot{x}(t)$ 
as the same order (small) quantities, cf. the remark after Eq. (\ref{H_eq}).
In passing let us remark that in our previous papers \cite{meie5,meie7,meie8} we have tentatively called (a)-(d) `the limit of general relativity' since under
these conditions the set of STG cosmological equations 
(\ref{00})-(\ref{matter_conservation}) reduces to those
of pure GR (with a cosmological constant if $V(\Psi_{\star})\neq 0$).

Subjecting the $\dot{H}$ equation (\ref{dotha}) to the approximation (\ref{xh_def}) gives in the first order
\be
{\dot H_{\star}} + {\dot h} = - \frac{3}{2} H_{\star}^2 - 3 H_{\star} h - \frac{1}{4\Psi_{\star}} \left( 1 + 
\frac {1}{2 A_{\star} \Psi_{\star}} \right) \frac{{\dot x}^2}{x} +
 \frac{1}{2 \Psi_{\star}} H_{\star} {\dot x} - \frac{3}{2} A_{\star} H_{\star}^2 x \,.
\label{dotHh}
\ee 
Taking the limit where the deviations $x, \dot{x}, h, \dot{h}$ vanish, we 
define 
\be
{\dot H_{\star}} = - \frac{3}{2} H_{\star}^2 \,,
\ee
which is familiar from the Friedmann solution of the dust dominated pure GR. 
It determines the time evolution of $H_{\star}$ to be 
\be \label{Hstar}
H_{\star} = \frac{2}{3(t - t_s)}\,.
\ee 
Here $t_s$ is a constant of integration which fixes the beginning of time scale; in what follows
we choose $t_s = 0$, $t > 0$.  
For late times when $H_{\star}$ is finite, Eq. (\ref{dotHh}) now assures that
$\dot{h}$ is also small, at least on a par with $h$.
To sum up, the appoximate first order equations read
\beq
 \label{ddotx}
{\ddot x} &=& \frac{{\dot x}^2}{2x} - 3H_{\star} {\dot x} + 3 A_{\star} \Psi_{\star} H_{\star}^2 x \,, \\
{\dot h} + 3 H_{\star} h &=& - \frac{1}{4\Psi_{\star}} \left( 1 + 
\frac {1}{2 A_{\star} \Psi_{\star}} \right) \frac{{\dot x}^2}{x} +
 \frac{1}{2 \Psi_{\star}} H_{\star} {\dot x} - \frac{3}{2} A_{\star} H_{\star}^2 x \,
\label{doth}
\eeq 
with $H_{\star}$ given by Eq. (\ref{Hstar}).
Notice that due to $H_{\star}$ the Eqs. (\ref{ddotx}), (\ref{doth}) 
depend explicitly on time $t$.
This means that the corresponding system of first order equations is not autonomous and the standard
phase space analysis is not applicable. However, we can straightforwardly integrate 
Eqs. (\ref{ddotx}), (\ref{doth}) in cosmological time and analyse the behaviour of solutions
in the neighbourhood of the limit of general relativity. 

For later reference let us note that the expansion (\ref{xh_def}) can be also applied for the effective barotropic index,
\be \label{weff}
{\rm w_{eff}} \equiv -1 - \frac{2 {\dot H}}{3 H^2} \approx 
- \frac{2}{3 H_{\star}^2} \left( {\dot h} + 3 H_{\star} h \right) \,.
\ee
Thus once $h(t)$ is found, it can be plugged into the equation above to reveal how $\rm w_{eff}$ evolves in cosmological time. In an analogous manner
one may also deal with $\frac{\dot{G}}{G}$ and other relevant quantities.


\section{Solutions in the cosmological time}

Despite its nonlinear and nonautonomous structure, one can solve Eq. (\ref{ddotx}) analytically.
It turns out that the type of the solution $x(t)$ depends on the constant 
\be
\label{D_def}
D \equiv 1 + \frac{8}{3} A_{\star} \Psi_{\star}
\ee
which characterizes the underlying STG. 
Then one can plug in $x(t)$ into Eq. (\ref{doth}) and solve the latter for $h(t)$, which also yields an analytic result.
Having found $x(t)$ and $h(t)$, it is possible to determine the evolution of the effective barotropic index ${\rm w_{eff}}$ from Eq. (\ref{weff}), and 
other quantities of interest.

\subsection{Polynomial solutions}

In the case $D > 0$ the  solution of Eq. (\ref{ddotx})  reads  
\be \label{x_exp}
\pm x(t) =  
 \frac{1}{t} \left( M_{1} t^{ \frac{\sqrt{D}}{2} }
- M_{2} t^{ -  \frac{\sqrt{D}}{2}} \right)^2 \,.
\ee
Here and below the ``$\pm$'' follows from an obvious 
invariance property of Eq. (\ref{ddotx}) 
under reflection $x \rightarrow -x$,  i.e. there are  solutions which lie in the regions 
$\Psi \gtrless \Psi_{\star}$ ($x \gtrless 0$), respectively.
The constants of integration $M_{1}$, $M_{2}$ are 
related to the initial data $x_*=x(t_*)$, $\dot{x}_*=\dot{x}(t_*)$ at some arbitrary time $t_*$ as
\beq
\label{M1}
M_1 &=& \frac{\dot{x}_* t_* + x_* (1+\sqrt{D})}{2 \sqrt{D}\sqrt{\pm x_*}} t_*^{\frac{1}{2}(1-\sqrt{D})} \,, \\
\label{M2}
M_2 &=& \frac{\dot{x}_* t_* + x_* (1-\sqrt{D})}{2 \sqrt{D}\sqrt{\pm x_*}} t_*^{\frac{1}{2}(1+\sqrt{D})} \,.
\eeq
Now we can integrate Eq. (\ref{doth}) to obtain
\be \label{h_exp}
\pm h(t) = \frac{2}{3 t^2} \left[M_{1}^2 \left(- a \sqrt{D} + b \right) t^{\sqrt{D}} 
+ M_{2}^2 \left(a \sqrt{D} + b \right) t^{-\sqrt{D}} + K  \right]  \,.
\ee
Here $K$ is another constant of integration and we have introduced 
constants $a$, $b$ which characterize the underlying STG,
\be \label{ab}
a \equiv \frac{3+6 A_{\star} \Psi_{\star}}{8 A_{\star} \Psi_{\star}^2} \,, \qquad
b \equiv \frac{3+10 A_{\star} \Psi_{\star}}{8 A_{\star} \Psi_{\star}^2}  \,.
\ee
As a result, the full Hubble parameter in the approximation under consideration is
\be 
H(t) = \frac{2}{3t} \left\lbrace 1 \pm \frac{1}{t}\left[ M_{1}^2 \left(- a \sqrt{D} + b \right) t^{\sqrt{D}} 
+ M_{2}^2 \left(a \sqrt{D} + b \right) t^{-\sqrt{D}} + K \right] \right\rbrace\,.
\ee
The effective barotropic index reads
\be \label{w_exp}
\pm {\rm w_{eff}}(t) = -\frac{\sqrt{D}}{t} \left[ M_{1}^2 \left(- a \sqrt{D} + b \right) t^{\sqrt{D}} 
- M_{2}^2 \left(a \sqrt{D} + b \right) t^{-\sqrt{D}} \right] \,.
\ee 

We can get a better feel of these solutions by considering their behavior
at certain limits and points. 
Asymptotically at $t \rightarrow \infty$ the solutions exhibit two distinct behaviors. For STGs with  $\sqrt{D} < 1$ (i.e. $A_{\star} \Psi_{\star} <0$) all
cosmological solutions irrespective
of their initial conditions monotonically approach   
  the general relativistic dust matter FLRW cosmology, 
 $\Psi (t)  \rightarrow \Psi_{\star} = {\rm const.}$, $H (t) \rightarrow  H_{\star} (t) =  2/(3t)$,
${\rm w_{eff}}(t) \rightarrow 0$, since all first order corrections vanish at this limit. 
On the other hand STGs with $\sqrt{D} > 1$ (i.e. $A_{\star} \Psi_{\star} >0$) allow only solutions that will diverge,
$x (t) \rightarrow \infty $, $h (t) \rightarrow \infty $, ${\rm w_{eff}}(t) \rightarrow \infty$, meaning that solutions in these theories can linger near general relativity only for a certain period, while as time evolves they will leave and the approximation scheme will break down eventually. 
(The case $\sqrt{D} = 1$ would imply $A_{\star}=0$ or $\Psi_{\star}=0$,
which contradicts the assumptions (c) or $0<\Psi<\infty$ of the present study.)

Taking $t \rightarrow 0$ the
quantities
$x(t)$ and $h(t)$ diverge for all integration constants and parameters of the theory except for 
the special $M_2=0$, $\sqrt{D} > 1$ case. This indicates that generally the solutions can not start near the limit of general relativity and only dynamical evolution can bring 
them close to it. 

The solution (\ref{x_exp}) also informs us that if the integration 
constants $M_1$ and $M_2$ are both positive or both negative, then 
at a finite moment 
\be
\label{t_b}
t_b = \left( \frac{M_2}{M_1} \right)^{\frac{1}{\sqrt{D}}} >0 
\ee
the corresponding solutions can go through 
\be
\pm x(t_b)=0 \,, \qquad \pm {\dot x}(t_b)=0 \,, \qquad \pm {\ddot x} (t_b) = 2D M_1 M_2 t_b^{-3} > 0 \,.
\ee
At $t_b$ these solutions do not stop at $x=0$, but bounce back, i.e. the solutions coming from the $x<0$ region return to the $x<0$ and similarly the solutions coming from  the $x>0$ region return to the $x>0$ region. There is no crossing from $x<0$ to $x>0$ or vice versa.
In terms of the initial data at some arbitrary time $t_*$ the bouncing solutions 
satisfy 
\be
\label{t_b_yes}
\mp \dot{x}_* t_* < \pm x_*(1-\sqrt{D}) \,, \quad \mathrm{or} \quad \pm \dot{x}_* t_* < \mp x_*(1+\sqrt{D}) \,, 
\ee
as can be inferred from Eqs. (\ref{M1})-(\ref{M2}).

In addition, at 
\be
\label{t_c}
t_c = \left( \frac{M_2 (1+\sqrt{D})}{M_1(1-\sqrt{D})} \right)^{\frac{1}{\sqrt{D}}} >0 \,,
\ee
the solutions may pass through
\be
\pm x(t_c)=\frac{4 D M_1 M_2}{(1-D)t_c} \,, \qquad \pm {\dot x}(t_c)=0 \,, \qquad \pm {\ddot x} (t_c) = - 2D M_1 M_2 t_c^{-3} \,.
\ee
This happens for two types of solutions: if $\mathrm{sign}(M_1) = \mathrm{sign}(M_2)$ for $\sqrt{D}<1$ and if $\mathrm{sign}(M_1) \neq \mathrm{sign}(M_2)$ for $\sqrt{D}>1$. 
The first type encompasses all $\sqrt{D}<1$ solutions which at $t_b$ have bounced back from 
$x=0$, now at $t_c > t_b$ they turn around again to proceed asymptotically towards $x=0$. The second type comprises of the $\sqrt{D}>1$ solutions which never
get to $x=0$, the moment $t_c$ marks their closest reach to $x=0$ before starting to flow away.

Therefore, in summary, the following picture emerges.
If $\sqrt{D}<1$ the solutions with initial conditions 
(\ref{t_b_yes}) first apporach $x=0$, then at $t_b$ reach $x=0$ and bounce back, further at $t_c$
turn towards $x=0$ again, to get there asymptotically as $t \rightarrow \infty$. The $\sqrt{D}<1$ solutions with
initial conditions outside the ranges given by (\ref{t_b_yes}) converge to $x=0$ monotonically.
If $\sqrt{D}>1$ the solutions with initial conditions 
(\ref{t_b_yes}) initially approach $x=0$, then at $t_b$ reach $x=0$ to bounce back and flow away. The $\sqrt{D}>1$ solutions with
initial conditions outside the ranges given by (\ref{t_b_yes}) move towards $x=0$, but before reaching it turn around at $t_c$ and leave. 
An exceptional case is the $\sqrt{D}>1$, $M_2=0$  solution which starts at $x=0$ and monotonically flows away from it.

As is evident from (\ref{h_exp}) and (\ref{w_exp})
the behavior of $h(t)$ and ${\rm w_{eff}}(t)$ is not
synchronous with $x(t)$. However, one can make some
simple generic observations
taking into account the definitions (\ref{D_def}), (\ref{ab}) and the basic assumption $0<\Psi<\infty$.
First, as $t \rightarrow 0$ the quantity ${\rm w_{eff}}(t) \rightarrow -\infty$ if $\frac{1}{9} < D < 1$, while ${\rm w_{eff}}(t) \rightarrow +\infty$ if $0 < D < \frac{1}{9}$ or 
$D >1$. 
Also, as we have noted above, $t \rightarrow \infty$ takes
${\rm w_{eff}}(t) \rightarrow 0$ if $D < 1$, while 
${\rm w_{eff}}(t) \rightarrow +\infty$ if $D > 1$.
Further, there can be specific moments
\be
t_d = \left( \frac{M^2_2}{M^2_1} \frac{(\sqrt{D}a + b)}{(-\sqrt{D}a + b)}\right)^{\frac{1}{2\sqrt{D}}}
\,, \qquad {\rm w_{eff}}(t_d)=0 \,, \qquad {\rm \dot{w}_{eff}}(t_d) \neq 0 
\ee
and
\be
t_e = \left( \frac{M^2_2}{M^2_1} \frac{(1+\sqrt{D}) (\sqrt{D}a + b)}{(1-\sqrt{D}) (-\sqrt{D}a + b)}\right)^{\frac{1}{2\sqrt{D}}}
\,, \qquad {\rm w_{eff}}(t_e) \neq 0 \,, \qquad {\rm \dot{w}_{eff}}(t_e) = 0 \,.
\ee
One can check that
$t_d>0$ if $0<D<\frac{1}{9}$, while $t_e>0$ if 
$0<D<\frac{1}{9}$ or $D>1$. 

Thus, the picture is the following. For $0<D<\frac{1}{9}$ the barotropic index ${\rm w_{eff}}(t)$ approaches ${\rm w_{eff}}=0$ from above, and passing this value at $t_d$, then later at 
$t_e$ the quantity ${\rm w_{eff}}(t)$ starts to increase again, and will asymptotically converge to the vanishing value. 
For $\frac{1}{9}<D<1$ the solutions exhibit a monotonic growth for ${\rm w_{eff}}$ which closes in to 
the ${\rm w_{eff}}=0$ dust matter regime from below.
The generic $D>1$ solutions start with decreasing 
${\rm w_{eff}}$, which reaches its lowest (and positive) value at $t_e$, but after that ${\rm w_{eff}}$ starts to increase again. The $D>1$, $M_2=0$ solution is an exception, here ${\rm w_{eff}}$ starts from $0$ and
keeps increasing in time.

\subsection{Logarithmic solutions}

In the case $D = 0$ ($A_{\star} \Psi_{\star} = - \frac{3}{8}$) the solutions 
of Eqs. (\ref{ddotx})--(\ref{weff}) read
\beq
\pm x(t) &=& \frac{1}{t} \left(\tilde{M}_{1} \ln t - \tilde{M}_{2} \right)^2 \,, \\
\pm h(t) &=& \frac{\tilde{M}_{1}}{3 \Psi_{\star} t^2} \left[  \frac{\tilde{M}_{1}}{2} (\ln t)^2 + (\tilde{M}_{1} - \tilde{M}_{2}) \ln t + \tilde{K} \right]\,,\\
\pm {\rm w_{eff}}(t) &=& - \frac{\tilde{M}_{1} }{2 \Psi_{\star} t} \left( \tilde{M}_{1} \ln t  + \tilde{M}_{1} - \tilde{M}_{2} \right) \,.
\eeq 
Here $\tilde{M}_{1}, \tilde{M}_{2} $ as well as $\tilde{K}$ are constants of integration, fixed by the initial data $x_* = x_*(\tilde{t}_*)$, $\dot{x}_* = \dot{x}_*(\tilde{t}_*)$ at some arbitrary time
$\tilde{t}_*$ as
\beq
\tilde{M_1} &=& \frac{\dot{x}_* \tilde{t}_* + x_*}{2 \sqrt{\pm x_*}} \, \tilde{t}_*^{\frac{1}{2}} \,, \\ 
\tilde{M_2} &=& \frac{(\dot{x}_*\tilde{t}_* + x_*)\ln \tilde{t}_* - 2x_*}{2 \sqrt{\pm x_*}} \, \tilde{t}_*^{\frac{1}{2}} \,.
\eeq

Unless $\tilde{M}_1=0$ the solutions exhibit the same generic behavior. They start by approaching $x=0$, at 
\be
\tilde{t}_b = \mathrm{e}^{\frac{\tilde{M}_2}{\tilde{M}_1}} 
\ee
reach
\be
\pm x(\tilde{t}_b)=0 \,, \qquad \pm {\dot x}(\tilde{t}_b)=0 \,, \qquad \pm {\ddot x} (\tilde{t}_b) = 2 \tilde{M}^2_1 \tilde{t}_b^{-3} > 0 \,
\ee
and bounce back, later at
\be
\tilde{t}_c = \mathrm{e}^{\frac{2\tilde{M}_1+\tilde{M}_2}{\tilde{M}_1}} > \tilde{t}_b 
\ee
the solutions pass through
\be
\pm x(\tilde{t}_c)=4 \tilde{M}^2_1 \tilde{t}_c^{-1} \,, \qquad \pm {\dot x}(\tilde{t}_c)=0 \,, \qquad \pm {\ddot x} (\tilde{t}_c) = - 2 \tilde{M}^2_1 \tilde{t}_c^{-3} \,,
\ee
and return flowing towards $x=0$ reaching it asymptotically in time. The effective barotropic index starts by decreasing from  a positive value, experiences 
\be
\tilde{t}_d = \mathrm{e}^{-\frac{\tilde{M}_1-\tilde{M}_2}{\tilde{M}_1}} \,, \qquad {\rm w_{eff}}(\tilde{t}_d) =0 \,, \qquad {\rm \dot{w}_{eff}}(\tilde{t}_d) = -\frac{\tilde{M}_1^2}{2 \Psi_{\star}} \tilde{t}_d^{-2}\\
\ee
but then starts to incease again at
\be
\tilde{t}_e = \mathrm{e}^{\frac{\tilde{M}_2}{\tilde{M}_1}} = \tilde{t}_b
\,, \qquad {\rm w_{eff}}(\tilde{t}_e) = -\frac{\tilde{M}_1^2}{2\Psi_{\star}} \tilde{t}_e^{-1} \,, \qquad {\rm \dot{w}_{eff}}(\tilde{t}_e) = 0 \,, \qquad {\rm \ddot{w}_{eff}}(\tilde{t}_e) = \frac{\tilde{M}_1^2}{2\Psi_{\star}} \tilde{t}_e^{-3}
\ee
reaching ${\rm w_{eff}}=0$ from below asymptotically in time.
In the $\tilde{M}_1 = 0$ case the solutions approach $x=0$ monotonically while ${\rm w_{eff}}$ is always zero.

\subsection{Oscillating solutions}

In the case $D<0$  solutions of Eq. (\ref{ddotx}) read
\be 
\pm x(t) = \frac{1}{t} \left[ N_{1} \sin \left(\frac{1}{2} \sqrt{|D|} \ln t \right) - 
N_{2} \cos \left(\frac{1}{2} \sqrt{|D|} \ln t \right) \right]^2 \,,
\ee
where the constants of integration $N_{1}$ and $N_{2}$ determined by the initial conditions $x_*=x(t_*)$, 
$\dot{x}_*=\dot{x}(t_*)$ at some arbitrary time $t_*$ as
\beq
N_1 &=& \frac{\sqrt{t_*}}{\sqrt{|D|} \sqrt{\pm x_*}} \left[
\left(\dot{x}_* t_* + x_* \right) \cos \left(\frac{1}{2} \sqrt{|D|} \ln t_* \right) +
x_* \sqrt{|D|} \sin \left(\frac{1}{2} \sqrt{|D|} \ln t_* \right)
 \right] \,, \\
N_2 &=& \frac{\sqrt{t_*}}{\sqrt{|D|} \sqrt{\pm x_*}} \left[
\left(\dot{x}_* t_* + x_* \right) \sin \left(\frac{1}{2} \sqrt{|D|} \ln t_* \right) -
x_* \sqrt{|D|} \cos \left(\frac{1}{2} \sqrt{|D|} \ln t_* \right)
 \right] \,.
\eeq
The corresponding solution of Eq. (\ref{doth}) is  
\beq 
\pm h(t) &=&  \frac{2}{3 t^2} \Bigg[ \left((N_{2}^2 - N_{1}^2) \frac{a  \sqrt{|D|} }{2} - N_{1} N_{2} \ b \right) 
\sin \left( \sqrt{|D|} \ln t \right)  \nonumber\\  
  &&\qquad + \left( (N_{2}^2 - N_{1}^2) \frac{b}{2} + N_{1} N_{2} \ a \sqrt{|D|}  \right)  
\cos \left(\sqrt{|D|} \ln t \right) + K \Bigg] \,,
\eeq
where $K$ is another constant of integration and $a$, $b$ are given by Eq. (\ref{ab}).
The full Hubble parameter now consists of small oscillations around the GR FLRW dust cosmological model 
\beq
H(t) &=& \frac{2}{3t} \Bigg\{ 1 \pm  \frac{1}{t} \Bigg[ \left((N_{2}^2 - N_{1}^2) \frac{a  \sqrt{|D|} }{2} - N_{1} N_{2} \ b \right) 
\sin \left( \sqrt{|D|} \ln t \right)  \nonumber\\  
  &&\qquad \qquad \qquad + \left( (N_{2}^2 - N_{1}^2) \frac{b}{2} + N_{1} N_{2} \ a \sqrt{|D|}  \right)  
\cos \left(\sqrt{|D|} \ln t \right) + K \Bigg] \Bigg\} \,.
\eeq
The effective barotropic index reads
\beq 
\pm {\rm w_{eff}}(t) &=& - \frac{\sqrt{|D|}}{t} \Bigg[ - \left( (N_{2}^2 - N_{1}^2) \frac{b}{2} + N_{1} N_{2} a \sqrt{|D|} \right) \sin \left(\sqrt{|D|} \ln t \right)  \nonumber
 \nonumber \\
 && \qquad \qquad \qquad + \left( (N_{2}^2 - N_{1}^2) \frac{a \sqrt{|D|}}{2} - N_{1} N_{2} b \right) \cos \left(\sqrt{|D|} \ln t \right)  \Bigg]  \,.
\eeq

The behavior of all these solutions is fairly simple as they approach the general relativistic dust matter cosmology 
in the manner of damped oscillations. At the moments
\be
t_b = \exp \left( \frac{2}{\sqrt{|D|}} \arctan \left(\frac{N_2}{N_1} \right) +\frac{2n \pi}{\sqrt{|D|}} \right)  
\ee
the deviation $x(t)$ of the scalar field passes through
\be
\pm x(t_b)=0 \,, \qquad \pm \dot{x}(t_b)=0 \,, \qquad 
\pm \ddot{x}(t_b) = \frac{-D}{2} (N_1^2+N_2^2) t_b^{-3} >0 \,,
\ee
i.e., bounces back from $x=0$, while at the moments
\be
t_c = \exp \left( \frac{2}{\sqrt{|D|}} \arctan \left( \frac{N_2 + N_1 \sqrt{|D|}}{N_1 - N_2 \sqrt{|D|}}\right) +\frac{2n \pi}{\sqrt{|D|}} \right)
\ee
it passes through
\be
\pm x(t_c) = \frac{-D}{1-D} (N_1^2+N_2^2) t_c^{-1} \,,
\qquad \pm \dot{x}(t_c) = 0 \,, \qquad
\pm \ddot{x}(t_c) = \frac{D}{2} (N_1^2+N_2^2) t_c^{-3} <0 \,,
\ee
i.e., turns around and evolves towards $x=0$ again. The amplitude of the deviations monotonically decreases while the period monotonically increases. The behavior of ${\rm w_{eff}}(t)$ is analogous, but not synchronous with $x(t)$. It is characterized by
\beq
t_d &=& \exp \left( \frac{1}{\sqrt{|D|}} \arctan \left(
\frac{(N_2^2-N_1^2)a \sqrt{|D|} - 2 N_1 N_2 b}{(N_2^2-N_1^2)b + 2 N_1 N_2 a \sqrt{|D|}} \right) + \frac{n \pi}{\sqrt{|D|}} \right) \,, \nonumber \\
{\rm w_{eff}}(t_d)&=&0 \,, \qquad {\rm \dot{w}_{eff}}(t_d) \neq 0 
\eeq
and
\beq
t_e &=& \exp \left( \frac{1}{\sqrt{|D|}} \arctan \left( -
\frac{(N_2^2-N_1^2)(b+a)D + 2 N_1 N_2 (b+aD)\sqrt{|D|}}{(N_2^2-N_1^2)(b+Da)\sqrt{|D|} - 2 N_1 N_2 (b+a) D} \right) + \frac{n \pi}{\sqrt{|D|}} \right) \,, \nonumber \\ 
\pm {\rm w_{eff}}(t_e) &=& \varsigma \frac{-D (N_2^2+N_1^2) \sqrt{b^2-a^2 D}}{2 \sqrt{1-D}} t_e^{-1}  \,,\qquad {\rm \dot{w}_{eff}}(t_e) = 0 \,, 
\eeq
where
\be
\varsigma = (-1)^n \, \mathrm{sign}\left( (N_2^2-N_1^2)(b+Da)\sqrt{|D|} - 2 N_1 N_2 (b+a) D \right) \,,
\ee
meaning oscillations around ${\rm w_{eff}}=0$ with exponentially decreasing 
amplitudes and exponetially increasing period.

\section{Discussion}
\subsection{Comparison with earlier results}

STG dust cosmology equations near the GR limit were investigated several years ago in the Einstein frame by 
Damour and Nordtvedt \cite{dn}. By invoking an analogy with 
a mechanical particle with time-dependent mass, they
demonstrated that in the case of coupling function 
$ (2 \omega (\Psi)+3)^{-1/2} \equiv \alpha (\varphi) = k \varphi $, $k= {\rm const.}$ the type of a solution 
 for the Einstein frame scalar field $\varphi (p) $ with the evolution parameter $p=(2/3) \ln t$
depends on the numerical value of the model-dependent constant $k$: the solution is exponential in time parameter $p$, i.e. 
polynomial in cosmological time $t$
if $0<k<3/8$, linear-exponential if $k = 3/8$ and oscillating if $k > 3/8$. 

In our earlier papers 
\cite{meie5} we investigated the Jordan frame scalar field equation close to the GR limit in 
the linearized approximation, found the fixed points and calculated the eigenvalues which determine
the type of solutions around these fixed points. Our results were qualitatively similar to 
those of  Damour and Nordtvedt \cite{dn}, but
the critical value of the model-dependent parameter turned out to be 3/16 instead of 3/8. 

In the present paper we refined the analysis and found solutions in the nonlinear approximation for 
the Jordan frame scalar field $\Psi(t)$ in 
the cosmological time $t$ and obtained the critical value of the model-dependent parameter
to be given by $A_{\star} \Psi_{\star} = - 3/8$.   
It is in exact agreement with the results of Damour and Nordtvedt, as the transformation 
between the Einstein and the Jordan frame quantities 
\be
(d \varphi)^2=\frac{2 \omega(\Psi) + 3}{4 \Psi^2} (d\Psi)^2
\ee
gives
\be 
k = \frac{d\alpha}{d\varphi}\big|_{\star} = \left[\frac{2 \Psi}{(2\omega + 3)^2} 
\frac{d\omega}{d\Psi}\right]_{\star} = -A_{\star} \Psi_{\star}\,.
\ee
It follows that the approximation used by Damour and Nordtvedt \cite{dn} in the Einstein frame
is congruent with our nonlinear approximation in the Jordan frame and thus can be considered
as an additional justification for our expansions (\ref{xh_def}),
(\ref{approx_2}).

\subsection{Combining the dust and potential dominated eras}

In the present paper we focussed upon the dust dominated cosmological epoch in the framework of 
STG with negligible scalar potential. 
In principle, this epoch could be followed by a scalar potential dominated epoch with insignificant
 matter density that we investigated by similar methods in our earlier papers \cite{meie7, meie8}. 
In both cases we assumed that the cosmological model has evolved towards the GR point $\Psi_{\star}$ (\ref{Psi_star})
since this is strongly indicated by different contemporary observations, and we solved field 
equations in a nonlinear approximation in the neighbourhood of this point. Let us now combine the 
conditions on the parameters of the models with the aim to view 
different epochs as parts of a single cosmological scenario.

In both cases there are general conditions for solutions to converge towards the GR value $\Psi_{\star}$ asymptotically in time: in the dust dominated model it reads (see Sec. 4)
\be
\label{dust attr}
A_{\star} \Psi_{\star} \equiv \left[ \frac{d}{d\Psi} \left( \frac{1}{2 \omega(\Psi) +3} \right) \Psi \right]_{\star} < 0
\ee
and in the potential dominated model \cite{meie7}
\be
\label{pot attr}
V (\Psi_{\star}) > 0 \,, \qquad \left[\frac{\Psi}{2V} \frac{dV}{d\Psi} \right]_{\star} < 1 \,.  
\ee
The converging solutions can be classified
according to the numerical value of a model-dependent parameter
as summarized in Table 1.
As discussed in Sec. 4, in the dust dominated epoch the behavior of the scalar field is determined by the quantity $D$ (\ref{D_def}) characterizing the STG model: the solutions are oscillating if $D<0$ ($A_{\star} \Psi_{\star} < - 3/8$), logarithmic 
if D=0 ($A_{\star} \Psi_{\star} = - 3/8$) and polynomial if 
$0<D<1$ ($ - 3/8 <A_{\star} \Psi_{\star} <0 $).   
In the scalar potential dominated models the corresponding classification can be given in terms of 
a model-dependent quantity 
\be 
\label{B_def}
B \equiv \left( A_{\star} \Psi_{\star} + \frac{3}{8} \right) -  A_{\star} \Psi_{\star}
\left[\frac{\Psi}{2V} \frac{dV}{d\Psi}\right]_{\star} \,
\ee
as follows: the solutions are oscillating in cosmological time if $B < 0$, linear-exponential
if $B = 0$ and exponential if $B > 0$ \cite{meie8}.
The same behavior carries over to the cosmological expansion 
as encoded in the Hubble parameter $H$ or barotropic index
$\mathrm{w_{eff}}$, i.e. polynomial, oscillating etc. convergence towards the dust FLRW values in the matter dominated epoch or de Sitter values in the potential dominated epoch, correspondingly.

A realistic STG cosmological scenario compatible with observations would better need to have GR as an attractor in
both dust dominated and matter dominated regimes.
Therefore for a credible STG 
both conditions (\ref{dust attr}) and (\ref{pot attr})
must be satisfied, thus constraining the set of functions 
$\omega(\Psi)$ and $V(\Psi)$ one can consider for constructing a viable model. The next filter is provided by qualitatively 
different behaviors among this converging class of models, e.g.
depending on $\omega(\Psi)$ and $V(\Psi)$ the evolution may 
be oscillating in the dust dominated and exponential in the
potential dominated epoch, etc., which might be possible to detect in future observations.

\begin{table}
\begin{center}
\begin{tabular}{l|lll} 
Epoch & & Solutions &
\\
\hline \hline \\
$\rho$ dominates & 
$\begin{array}{l} \mathrm{oscillating} \\ D < 0 \vspace{0.1cm} \end{array}$ &
$\begin{array}{l} \mathrm{logarithmic} \\ D = 0 \vspace{0.1cm} \end{array}$ & 
$\begin{array}{l} \mathrm{polynomial} \\  0 < D < 1 \vspace{0.1cm} \end{array}$ 
\\
\hline  \\
$V$ dominates &
$\begin{array}{l} \mathrm{oscillating} \\ B<0 \end{array}$ &
$\begin{array}{l} \mathrm{linear-exponential} \\ B=0 \end{array}$ &
$\begin{array}{l} \mathrm{exponential} \\ 0<B \end{array}$ 
\\
\end{tabular}
\end{center}
\caption{Classification of the qualitative behavior of solutions of the scalar field and cosmological expansion while converging to the GR limit in the dust matter ($\rho$) dominated and potential ($V$) dominated epochs, determined by the parameters $D$ (\ref{D_def}) and $B$ (\ref{B_def}), which characterize the underlying STG.}
\end{table}

\section{Summary and outlook}
In this paper we have considered generic Jordan frame STG flat FLRW cosmological models in the dust dominated era with negligible scalar potential near the limit of general relativity as favored by 
various observational constraints. We derived and solved nonlinear approximate equations for small deviations of the scalar field and cosmological expansion from their GR limit values.
Depending on the scalar field coupling function $\omega(\Psi)$ the models
fall into two classes where either all solutions approach GR asymptotically in time or only a single fine-tuned solution does. The models with universally converging solutions come in three characteristic types: 
polynomial convergence, logarithmic convergence, and 
damped oscillations around general relativity.

The approximation scheme assumes that the first derivative of $\frac{1}{2\omega(\Psi)+3}$ w.r.t. $\Psi$ evaluated at
the GR limit (\ref{Psi_star}) is nonvanishing and finite, while the higher dervatives do not diverge. Then the only parameter
characterizing the 
underlying distinct STG which enters the approximation equations and the  analytic solutions is the value of the first derivative.
Thus in principle the present study
encompasses a very large generic family of STG models
and in this sense has wider applicability than considering
example models, equvalent to a particular form of $\omega(\Psi)$ chosen.

The class of STGs where the GR limit is an attractor for the nearby solutions is of interest
because there is a dynamical mechanism naturally driving the solutions to satisfy observational constraints.
So, combining the results of the present work on the dust dominated
epoch with earlier results on the potential dominated regime, 
provides a reasonable viability filter for STG models 
in terms of the conditions (\ref{dust attr}), (\ref{pot attr}).

On the other hand the converging solutions still have their
characteristic small deviations from the ruling $\Lambda$CDM scenario. Given the generic analytic solution for the cosmological expansion and the corresponding effective barotropic index near the GR limit,
it remains as a future work to face it with actual data
and to draw observational constraints on the STG models.
Similarly, the expansion history enters as background evolution
in the equations for the growth of perturbations, which leads to 
another line of investigation. Finally, the dust and late-time potential dominated epochs must be patched together with an account of the early universe.

\bigskip
{\bf Acknowledgements}

This work was supported by the Estonian Science Foundation
Grant No. 8837 and by Estonian Ministry for Education and Science
Support Grant No. SF0180013s07. 

\end{document}